# The mathematical representation of physical objects and relativistic Quantum Mechanics.

Enrique Ordaz Romay[1]

*Facultad de Ciencias Físicas, Universidad Complutense de Madrid*

## Abstract

The mathematical representation of the physical objects determines which mathematical branch will be applied during the physical analysis in the systems studied. The difference among non-quantum physics, like classic or relativistic physics, and quantum physics, especially in quantum field theory, is nothing else than the difference between the mathematics that is used on both branches of the physics. A common physical and mathematical origin for the analysis of the different systems brings both forms (quantum and classic) of understanding the nature mechanisms closer to each other.

---

[1] eorgazro@cofis.es

# Introduction

Inside any physical system there are groups of elements that are usually considered as individual units that interact with each other, but in whose interior other forces intervene that are different from those that are usually considered at the level of the system as a whole. Accordingly, for example, when we study the interactions among solids we usually don't keep in mind the intermolecular forces or when we study the gravitational interactions among planets we do not take into account the geologic planetary forces.

Usually, this simplification is carried out using transformations (simplifications that lead us to symmetries and changes of coordinates based on such symmetries) that allow us to obtain later on, simple and convenient expressions and solutions for our systems under study.

To carry out these transformations (simplifications plus changes of coordinates) we should "make a model" of the physical objects we work with, by means of an appropriate mathematical representation. We call these transformations that allow us to make a model of an object, the mathematical representation of the physical object. This way, our physical system will be formed by a set of equations that represent the real physical objects.

In quantum physics, the physical objects are the particles of the system and the mathematical representation of these physical objects is the wave function. However, a narrow relationship exists between the wave function and the action function of the system, since both of them are maximal representations of the system. This observation allows us to obtain, from the classic analysis (relativistic) perspective, the complete development of the quantum theory of fields.

# Physical systems.

A physical system is defined by a delimited region   of 4-space. All the groups of points $X = (x^0, x^1, x^2, x^3)$ compose this region that we consider by definition to conform the system.

A physical magnitude is a degree 3 or 4 scalar, vector, matrix or tensor field that is defined inside the system[2]. When the field is a scalar, this magnitude $m(X)$ is called mass; when the field is a vector, this magnitude is called 4-potential vector $e$ $A(X)$; …

Let $P(X)$ be the linear momentum due to a group of magnitudes. That is to say, for scalar and vector magnitudes [1], the form of the linear momentum is:

$$P(X) = m(X)\, v + e A(X)$$

On the other hand, a physical system is defined by a function that we call action $S$ (traditionally scalar, although of tensor origin [2], at the moment we will not look at its degree) that describes the system at its maximal way. That is to say, if $S$ represents a system defined in the   region and $P$, as linear momentum, represents the magnitudes, then [3] (the negative sign only is traditional):

$$S = -\int_\Omega P \qquad (1)$$

This equation comes to say that, *the complete definition of a physical system, to which we call action, is defined by the space-time evolution of the physical magnitudes that the system contains represented by the linear momentum.*

If we reverse the expression (1) we obtain:

$$P_i = -\frac{\partial S}{\partial x^i} \qquad (2)$$

---

[2] The maximum degree that a field could have in a space of four dimensions is four (Refer to [1] physics/0306134 *The action as a differential n-form and the analytic deduction of the nuclear potentials* ).

This expression coincides with the four-impulse in classic theory of fields [4], when the action is expressed as a function of the coordinates. This second equation says that, *the physical magnitudes of a system are the variations or non-homogeneities that there are in the system in respect to their four-coordinates.*

## Mathematical representation of physical objects.

Usually, the expressions of physical systems use non-Cartesian coordinate systems. These coordinates are called generalized coordinates [5]. The reason for the coordinates change is to obtain simpler expressions for the system equations. We get this observing the symmetries.

A coordinate change implies the election of some equations $\Psi^i = \Psi^i(x^0, x^1,...)$ that applied over the sets $U_i$ ($\Omega \subset \bigcup_i U_i$) conform a mapping on Ù. Substituting in (2) we obtain: $P_i = -\dfrac{\partial S}{\partial \Psi^k}\dfrac{\partial \Psi^k}{\partial x^i}$.

Using the tensor form of the action [1] of range one, we find:

$$P_i^j = -\frac{\partial S^j}{\partial \Psi^k}\frac{\partial \Psi^k}{\partial x^i} \qquad (3)$$

From the expression (3) we deduce that, at first, there are two ways to understand the physical form of the Ø coordinates:

- Knowing the relationship $\dfrac{\partial \Psi^k}{\partial x^i}$, just as we have made it at the beginning.
- Knowing $\dfrac{\partial S^j}{\partial \Psi^k}$.

The first case is the traditional one. In this case the geometric symmetries induce a change of coordinates that simplifies the equations of the system. The cylindrical, spherical and hyperbolic coordinate transformations belong to this type.

In the second case we can simplify the equations of the system if we look for a transformation that is simple from the action point of view.

Let our system be formed by physical objects in such a way that, *P* takes positive values inside the physical objects, while its value is zero outside of such objects. In this case, the function *P(x)* for each point of Ù, turns into a distribution function for the magnitudes that the system contains at that point.

In line with the expression (3) the linear momentum *P* has the form *P(x)* = *P(Ø(x))*. Since *P* characterizes the physical magnitude, the function *Ø(x)* represents a distribution function that characterizes the object or objects of the system's magnitude. That is to say: *in the transformation that simplifies the expression* $\frac{\partial S^j}{\partial \Psi^k}$, *the function Ø(x) is the mathematical representation of the system's physical objects.*

## Quantum theory of fields.

In quantum theory the objects (particles) come to be represented by a function Ø that is called wave function. The square of this function is in fact a distribution function.

The wave function of a system is a maximal representation of the system, same as the action function is. That is to say that both, the action *S* and the wave function Ø represent the same physical concept, with the difference that the action has units of power, while the wave function lacks of units.

An important property in quantum field theory is that the wave function has a tensorial character. This fact, far from being a restriction, actually it is a generalization of the concept of wave function in classic quantum mechanics, besides being the natural form of deducing the spin concept.

This way, the degree of the wave function tensor speaks to us of the spin of the system [6]. A system of zero spin will be defined by a scalar function, that is to say, it is a tensor wave function of zero degree (without indexes). A spin system ½ will come defined by a vector function, that is to say, a degree-one tensor wave function (with a single index). A spin one system will be defined by a matrix function, that is to say, a degree-two tensor wave function (with two indexes).... For the sake of simplicity in the expressions we will use vector notation, keeping into account that this restriction in the notation does not suppose a restriction in the validity of our expressions whose generalization to a system of generic spin is trivial.

The simplest form of relating the action $S$ and the wave function $\emptyset$ would be making $\frac{\partial S^j}{\partial \Psi^k} = cte$, meaning a lineal relationship $S^j \sim \emptyset^j$. The proportionality constant will depend on the relationship existing between the system and the wave function. Such differences are the properties that we observe in quantum particles that characterize them against the traditional physical systems. These properties are:

1. The wave function represents a very small object compared to a traditional physical system; therefore we will use a small proportionality constant having units of energy, which is represented by $\hbar$.
2. The particles are characterized by an oscillating behaviour compared to the systems that are traditionally considered static. Therefore, such a constant should contain the imaginary factor $i = \sqrt{-1}$.

Substituting these two considerations in the relationship between S and Ø, we obtain [1]:

$$S^j = i\hbar \Psi^j \quad (4)$$

The proportionality constant $\hbar$ has been named the Planck constant [7], it has units of power and it is dependent of the units system that we use. In international units system $\hbar = 1.0545 \cdot 10^{-34}$ J s

The expression (4) tells us that *the wave function represents a quantum system as a whole, with the particularity that the particles, compared to our traditional*

*macroscopic observations, have a lower magnitude, equivalent to the Planck constant and a vibratory behaviour.*

We also see that the expression (4) forces us to consider the action as a tensorial object whose degree is equivalent to that of the wave function of the system. This consideration, same as in the case of the wave function, it is a generalization of the particular case of the scalar action seen in traditional classic mechanics. This is even more justified than in the equation (3) of the general form of fields theory, whereas in quantum theory this tensor expression has a concrete physical reality: same as the tensorial wave function, the tensorial action represents the spin of the system.

The method for passing from a tensor representation to a scalar one is performing the corresponding contraction [2]: $S = S_i^i = \sqrt{S^i S_i}$. Applying the property of tensors whereby the product of two tensors is constant and independent of the change in reference system and keeping in mind that a variation in the action when this is a function of the coordinates is really a coordinates transform, we deduce that $\delta S = \delta S_i^i = 0$; mathematical expression of the principle of least action.

# Mathematical representation of physical objects in quantum theory of fields.

The form by which the contained objects in a system are represented in quantum field theory is by means of the wave function. This way, the existent relationship among the mathematical representation of the quantum physical object (that is frequently called particles assembly) and the system is:

1. The assemblies of particles represent very small objects in comparison with the usual units of the macroscopic systems.
2. The assemblies of particles are characterized by a vibratory behaviour, compared to the classic systems in which, in the fundamental state of energy the physical objects remain at rest.

These properties, as we already saw, lead to the equation (4). If we combine this expression with equation (3) generic for field theory, we deduce:

$$P^{ijk\ldots}(\Psi^{jk\ldots}) = -i\hbar \frac{\partial \Psi^{jk\ldots}}{\partial x_i} \quad (5)$$

These equations represent the canonical quantization.

From the traditional quantum point of view, the expression (5) represents the duality between the representation of the particle as a corpuscle in the first term of the expression and the particle as a wave in the second term. The particularization for linear vector momentums, that is to say, with a single index, leads to the traditionally well-known canonical quantization [8].

The step from the expression (5) to the transformations of canonical quantization of non-relativistic quantum mechanics requires two additional considerations: On one hand, the linear momentum in the non relativistic version is deduced from the physical-geometric macroscopic conditions and not starting from the wave function. On the other hand the wave function does not represent to the complete system but only the particle that is being studied, considering the rest of the system as a macroscopic object. The result of these two considerations becomes the transformation of the expression (5) into the well-known transformations of canonical quantization. If we express them in components and making $c\ P^0 = -H$ being $c$ the light speed and $H$ the Hamiltonian of the system, we obtain:

$$P_x(x,y,z,t)\cdot\phi(x,y,z,t) = -i\hbar\frac{\partial\phi}{\partial x}, \quad P_y(x,y,z,t)\cdot\phi(x,y,z,t) = -i\hbar\frac{\partial\phi}{\partial y}$$
$$P_z(x,y,z,t)\cdot\phi(x,y,z,t) = -i\hbar\frac{\partial\phi}{\partial z}, \quad H(x,y,z,t)\cdot\phi(x,y,z,t) = i\hbar\frac{\partial\phi}{\partial t} \quad (6)$$

When we introduce here as wave function for non relativist mechanics the function ø instead of Ø we want to emphasize the difference among the non relativist wave function that represents a single particle inside the system and the relativistic one which represents the assembly of particles that constitute the complete system.

# Historical reflection.

The development of the quantum and relativity theories took place in a contemporary way during the first years of the XX century. This way, the expression (2), for any type of metric could not be deduced in a general way before 1915 (formulation of the General Relativity), while the expressions (6) became known from 1923 (formulation of the equation of Schrödinger). This small time difference caused that both theories were developed in an independent way, just as if did not exist any relationship among them.

However, if the expression (2) had been deduced long before, the reasoning that leads us from this equation to the expression (6) would have become "natural", in lack of knowing the value of the Planck constant (just as it happened with the law of Avogadro in which the value of the constant was discovered much later).

Regarding the expression (4), the relation between $S$ and $Ø$, where the action is a magnitude related with the space-time, while the wave function represents the matter, can only be deduced in a natural way considering the duality between both concepts. If the space is represented in a real way, the matter should be represented in a non-real way, that is to say imaginary.

This form of reasoning is completely similar to the one used by Fermat while he was establishing the geometric optics in which the iconic (?) function $y$ representing the properties of the light beam is related with the matter by means of a relation that involves the imaginary constant: $f = ae^{iy}$ [4].

# Conclusion.

From a mathematical point of view the analysis of a physical system supposes to choose an expression for each physical magnitude and to relate such expressions to obtain the equations that describe the system to us.

The expressions that are used for the physical magnitudes will determine the mathematics' branch that will be applied and therefore the interpretive apparatus that will allow to translate the mathematical results into physical results.

Consequently, the unification of the different branches of physics, especially the classic and quantum physics, should be focused on the search for common and general mathematical representations which simplifications, in function of the properties of the systems, are the source of the present diversity of physical theories.